\documentstyle[prl,aps]{revtex}

\begin{document}

\preprint{\vbox{ \hfill SNUTP-97-068 }}
\title{ Vector  mesons in medium with finite three momentum.}
\author{Su Houng Lee }
\address{Department of Physics, Yonsei University, Seoul, 120-749, Korea}
\date{\today}
\maketitle
\begin{abstract}
We formulate a QCD sum rule to find the three momentum 
 dependence of the peak position of the vector  meson spectral density 
in nuclear medium.     
 We find less than 2 \% (0.1 \%) shift of the peak position at nuclear matter density and at (${\bf q} =0.5$GeV/c) for the $\rho,\omega$ ($\phi$) mesons.  
However, at higher density  and three momentum, its effect becomes non-negligible and  the relevance of our result in the dilepton spectrum in A-A and p-A reaction is discussed.  

\end{abstract}


The properties of vector mesons in nuclear medium have attracted a lot of 
interest because of their potential role to experimentally observe
a non-perturbative aspect of QCD, namely the restoration of  spontaneously 
broken chiral symmetry at finite temperature or density, through dileptons from
  A-A or p-A reaction\cite{Hatrev}.

Many model calculations have been performed to calculate the vector meson mass 
shift at finite density.  By now,  there seems to be a consensus that the 
average  peak position of the vector meson spectral density at zero three momentum (${\bf q}=0$) will shift down at finite density\cite{BR91,HL92,Weise97,Friman96,Walecka}.

Indeed, there already exist dilepton data from the CERES collaboration,
which report an enhancement of low mass dileptons below the $\rho$ meson 
invariant mass in  S-S and recently from Pb-Pb collisions at CERN
\cite{Ceres}.  
So far, all conventional collision models failed to explain the enhancement 
except when the $\rho$ meson mass is allowed to decrease in medium as predicted
by theoretical calculations\cite{Ko}.  
However, before coming to any definite conclusions, it is  
necessary to consider all possible conventional mechanisms.

In nuclear medium, in addition to possible change in the vector meson mass, 
there will be breaking of Lorentz invariance and hence two independent
polarization directions of the vector mesons.  Each polarization will have 
a different dispersion relation, which  to leading order in the 
three momentum would be modified 
to $\omega^2-(1+a) {\bf q}^2-m_V^2=0$ with  $a \neq 0$.   
  Suppose, experimentally, one detects a dilepton with energy $\omega$ and three
momentum ${\bf q}$, then the average peak due to the vector meson will appear at
  $M^2=(m_V^2+a {\bf q}^2)$, so that if $a <0$, the strength will be shifted
downwards for dilepton pairs with  ${\bf q} \neq 0$.  
Hence, it is important to estimate the finite ${\bf q}$  effect.  

In this letter, we  formulate a QCD sum rule to find the finite ${\bf q}$ effect
 to leading order in density
  for the transverse and longitudinal direction of the $\rho,\omega$ and $\phi$ mesons.  
This formalism also provides the first attempt to 
estimate the leading ${\bf q}$ dependence of the $V-N$ T matrix, when it has a 
 small off-shell dependence.   
 
Let us consider the correlation function between the  vector  current with 
 $\rho$, $\omega$ and $\phi$ meson quantum numbers, 
 $J_\mu^{\rho,\omega}=\frac{1}{2}(\bar{u} \gamma_\mu u  \mp 
\bar{d} \gamma_\mu d)$ and $J_\mu^\phi=\bar{s} \gamma_\mu s$  
in the nuclear medium. 

\begin{eqnarray}
\Pi_{\mu\nu} (\omega,{\bf q}) &=& i \int d^4x e^{iqx}\langle 
      T [ J_\mu(x) J_\nu(0) ] \rangle_{n.m.}.
\label{pol1}
\end{eqnarray}

Here $ \langle \rangle_{n.m.}$ denotes the nuclear matter 
expectation value.
In general, because the vector current is conserved, 
 the correlation function in  eq.(\ref{pol1}) will have  two invariant
 functions\cite{tensors}. 

\begin{eqnarray}
\Pi_{\mu\nu}(\omega,{\bf q})=\Pi_T q^2 {\rm P}^T_{\mu\nu}+ \Pi_L q^2 
 {\rm P}^L_{\mu\nu},
\label{pol2}
\end{eqnarray}   
where for $q=(\omega,{\bf q})$ and medium at rest, we have, 
$ 
{\rm P}^T_{00}  =  {\rm P}^T_{0i}={\rm P}^T_{i0}=0,~~
{\rm P}^T_{ij}  =  \delta_{ij}-{\bf q}_i {\bf q}_j/{\bf q}^2,$ and $
{\rm P}^L_{\mu\nu}  =  (q_\mu q_\nu/q^2-g_{\mu\nu}- {\rm P}^T_{\mu\nu}) $.
In the limit when ${\bf q}\rightarrow 0$, there is only one invariant function
$\Pi_L(\omega,0)=\Pi_T(\omega,0)$.  
In this work, 
we will formulate a QCD sum rules for both $\Pi_L(\omega,{\bf q})$ and  
 $\Pi_T(\omega, {\bf q})$ at finite ${\bf q}$.   

The starting point is the 
energy dispersion relation at finite ${\bf q}$.  For small ${\bf q}^2<\omega^2
$, we 
can make a Taylor expansion of the correlation function such that,
 \begin{eqnarray}
{\rm Re} \Pi_{L,T}(\omega^2,{\bf q}^2)  =  {\rm Re} \left( \Pi_{L,T}^0(\omega^2,0)+ 
\Pi_{L,T}^1(\omega^2,0) ~ {\bf q}^2 + \cdot \cdot \right) \nonumber \\  
  =  \int_0^\infty du^2 \left(
 {\rho(u,0)_{L,T}^0 \over (u^2-\omega^2)} +  {\rho(u,0)_{L,T}^1 \over (u^2-\omega^2)} ~ {\bf q}^2 + \cdot \cdot \right) , 
\label{dis1}
\end{eqnarray}
here $\rho(u,{\bf q})=1/\pi {\rm Im} \Pi^R(u ,{\bf q})$,  and 
$R$ denotes the retarded correlation function.  
As we will discuss below, the left hand side (${\rm Re} \Pi$) 
is known  only to leading order
in ${\bf q}^2$ and since we are interested in the leading  behavior anyways,
we will look at  the  dispersion relation for $\Pi^1$ in eq.(\ref{dis1}).

The real part of eq.(\ref{dis1})  is calculated via 
the Operator Product Expansion (OPE) at large $-\omega^2 \rightarrow 
\infty$ with  finite ${\bf q}$.  The full polarization tensor will have the following form.
\begin{eqnarray}
\Pi_{\mu \nu}(\omega,{\bf q})    =  & 
(q_\mu q_\nu-g_{\mu \nu}) 
\left[- c_0 {\rm ln}|Q^2|+ \sum_d {c_{d,d} \over Q^d} A^{d,d}(n.m.) \right] 
\nonumber \\[12pt]
 + &  \sum_{s,\tau=2} {1 \over Q^{s+\tau-2}} 
\left[ c_{d,\tau} 
g_{\mu \nu}  q^{\mu_1} \cdot \cdot q^{\mu_s} A^{d,\tau}_{\mu_1 ..\mu_s}
(n.m.) + \cdot \cdot  \right],
\label{ope1}
\end{eqnarray}
where, $Q^2={\bf q}^2-\omega^2$. 
Here, $A^{d,\tau}(n.m.)$ represents the nuclear matter expectation 
value of an operator of 
 dimension $d$ and twist $\tau=d-s$, where $s$ is the number of spin 
index.  These operators are defined at the 
scale  $Q^2$ and the $c$'s are the 
dimensionless Wilson coefficients  with the running coupling constant.   
This way of including the density 
effect is consistent at low energy\cite{HKL93}. 
The first set of terms in eq.(\ref{ope1}) 
come from the OPE of 
scalar operators, the second set from  operators with non zero spin $s$.

To linear order in density, the matrix elements are related to the nucleon
expectation values of the operator via 
\begin{eqnarray}
A(n.m.)= A_0+ n_n A_N,
\end{eqnarray}
where $A_0$ is the corresponding vacuum 
expectation value, $A_N$ the nucleon expectation value and $n_n$ 
the nuclear density.  

As in the vacuum, we will truncate our OPE up to dimension 6 operators.  This  
implies that in our OPE in eq.(\ref{ope1}), we will have contributions from 
 $(\tau,s)=(2,2),(2,4),(4,2)$.   The nucleon 
matrix elements of the $\tau=2$ operators are 
well known.  The $\tau=4$ matrix element appearing in the $\rho,\omega$ sum rule are similar to those appearing in electron DIS\cite{Jaffe} and have been estimated  \cite{CHKL,L94}
up to about $\pm$ 30\% uncertainty from 
available DIS data from CERN and Slac.  

 The ${\bf q}$ dependence coming from the first line of  eq.(\ref{ope1}),
namely the contribution from the scalar operators,  come from the  ${\bf q}$
dependence in  $Q^2$.
These form the so called ``trivial'' ${\bf q}$ dependence, 
and comes from  replacing $\omega^2 \rightarrow \omega^2-{\bf q}^2$ when 
going from zero to finite three momentum.  
Here, we are not interested in these trivial dependence  and also not in the 
possible change in the scalar mass $m_V$.  Consequently we do not
need the nucleon expectation value of the scalar operators. Operators with spin 
also partly contribute to the change in the scalar mass
and hence also to the trivial changes.  However, these spin parts  also
give the non-trivial ${\bf q}$ dependence.   
 A  prescription to find the nontrivial ${\bf q}^2$ 
dependence in the  OPE is to first calculate the total ${\bf q}^2$ term 
 $\Pi^1$ and 
then subtract out the trivial dependence.

\begin{eqnarray}
\frac{1}{\omega^n}(1+ d \frac{{\bf q}^2}{\omega^2})
~~\rightarrow (d-\frac{n}{2})
\frac{{\bf q}^2}{\omega^{n+2}}.
\label{pres}
\end{eqnarray}

 Following this prescription, we find the following contributions from 
the $\tau=2,4$ operators.  

\begin{eqnarray}
\Pi_{L,T}^1(\omega)/\rho_n= {b_2 \over \omega^6} + {b_3 \over \omega^8}.
\label{ope2}
\end{eqnarray}

 For $\rho, \omega$, the transverse  (T) and longitudinal (L) parts give,  
\begin{eqnarray}
b_2^T & =& (\frac{1}{2}C_{2,2}^q -\frac{1}{2}C_{L,2}^q) mA_2^{u+d}  + 
(C_{2,2}^G -C_{L,2}^G) mA_2^{G}, \\ 
b_3^T & = & (\frac{9}{4}C_{2,4}^q -\frac{5}{2}C_{L,4}^q) m^3A_4^{u+d}  + 
(\frac{9}{2}C_{2,4}^G -5C_{L,4}^G) m^3A_4^{G}  + \frac{1}{2} m \left(
-(1+\beta)(K^1+\frac{3}{8}K^2+\frac{7}{16}K^g)+K^1_{ud}(1\pm1)
\right),  \\
b_2^L & =&  -\frac{1}{2}C_{L,2}^q mA_2^{u+d}  -C_{L,2}^G mA_2^{G}, \\ 
b_3^L & = & (\frac{1}{2}C_{2,4}^q -\frac{5}{2}C_{L,4}^q) m^3A_4^{u+d}  + 
(C_{2,4}^G -5C_{L,4}^G) m^3A_4^{G} + \frac{m}{8}(1+\beta) 
\left( K^2-\frac{3}{2}K^g \right),
\label{pil}
\end{eqnarray} 
where $\pm$ refers to  the $\rho$ and $\omega$ case.  
Here, $m$ is the nucleon mass and for even $n$,  
\begin{eqnarray}
A_n^q & = &2 \int_0^1 dx x^{n-1} [ q(x,Q^2)+ \bar{q}(x,Q^2)],  \nonumber \\
A_n^G & = & 
2 \int_0^1 dx x^{n-1} G(x,Q^2),
\end{eqnarray}
where $q(x,Q^2)$ and $G(x,Q^2)$ are the quark and gluon distribution functions. 
We will use the HO parameterization for these obtained in ref\cite{Reya92} which
 should be used 
with the Wilson coefficients $C's$ in the $\bar{\rm MS}$ scheme\cite{Muta78}.  
 Terms proportional to $K's$ come from $\tau=4,s=2$.   We will use the set of
 K values obtained in ref.\cite{CHKL}; ( $K^1,K^2,K^1_{ud},K^g)=(-0.173{\rm GeV}^2,0.203{\rm GeV}^2,
-0.083{\rm GeV}^2,-0.238{\rm GeV}^2)$ and 
we take $\beta=0.5$.  For the $\tau=4$ operators, we neglect the $Q^2$ 
dependence.  

For the $\phi$ meson,
\begin{eqnarray}
b_2^T & =& (2C_{2,2}^q -2C_{L,2}^q) mA_2^{s}  + 
(2C_{2,2}^G -2C_{L,2}^G) mA_2^{G} \\ 
b_3^T & = & (9C_{2,4}^q -10C_{L,4}^q) m^3A_4^{s}  + 
(9C_{2,4}^G -10C_{L,4}^G) m^3A_4^{G}-\frac{5}{2}mC,\\ 
b_2^L & =&  -2C_{L,2}^q mA_2^{s}  -2C_{L,2}^G mA_2^{G} \\ 
b_3^L & = & (2C_{2,4}^q -10C_{L,4}^q) m^3A_4^{s}  + 
(2C_{2,4}^G -10C_{L,4}^G) m^3A_4^{G}-9mC,
\end{eqnarray}
where $C$ is defined from $\langle N(p)| \bar{s} D_\mu D_\nu m_s s | N(p)
 \rangle  =(p_\mu p_\nu -\frac{1}{4} m^2 g_{\mu \nu}) C$.  We let $C =-x_s \langle N| \bar{s} m_s s | N \rangle =-0.1548 \cdot x_s$GeV$^2$.  Where we used 
the same 
number for the strange content of the nucleon as in ref.\cite{Asa}  with the 
 normalization of $\langle p | p \rangle =2 \omega (2 \pi)^3 \delta^3(p-p)$. 
 $x_s$ is an unknown number, however, assuming its order to be similar to the 
ratio $A^s_4/A^s_2$, we  will take $x_s=0.04$.  

Table 1 summarizes each contribution to the $b$ coefficients at $Q^2=1$GeV$^2$.
As can be seen from the table, for the transverse part of the $\rho,\omega$, 
 the well known twist-2 quark contribution dominates 
both the $b_2,b_3$.  However, for the longitudinal part,
both the quark and gluon twist-2 part contributes at order $\alpha_s$ and  the 
twist-4
part becomes important, which has larger uncertainty.  This is also roughly true
for the $\phi$ meson.  
At higher $Q^2$ values, the values of $b$'s decreases in general. 

In the vacuum, the spectral density appearing in the left hand side of 
eq.(\ref{dis1}) is modeled with a pole and a continuum:
 $8 \pi^2 \rho(u)=F \delta (u^2-m_V^2) +c_0 \theta(u^2-S_0) $.   
 In our case, we
allow the three parameters to vary non-trivially by a term proportional to 
 ${\bf q}^2$ and 
the nuclear density $n_n$.  Such as, $ F \rightarrow F+f \cdot {\bf q}^2,~~
m_V^2 \rightarrow m_V^2+a \cdot {\bf q}^2,~~
S_0 \rightarrow S_0+s \cdot {\bf q}^2$ .  
Also, when using the dispersion relation in eq.(3), there 
is a possible ambiguity in the subtraction constants and we have to
know if there exist singularities in the spectral density of the form
 $\delta(u^2)$ or $\delta'(u^2)$.  The contribution proportional to $\delta(u^2)$ is not known.  However, the other singularity can be unambiguously calculated from the the nucleon-hole contribution.  This term is called the scattering
term\cite{HL92} and gives a non-trivial contribution to the longitudinal direction.  Hence, the  spectral density of $\Pi^1(\omega)$ is,
\begin{eqnarray}
8\pi^2 \rho^1(u) & = & f \cdot \delta(u^2-m_V^2)-a \cdot F \delta'(u^2-m_V^2)
 \\  && -s \cdot c_0 \delta(u^2-S_0)+8 \pi^2 n_n b_{scatt} \delta'(u^2),
\label{phen2}
\end{eqnarray}
where, $b_{scatt}= 1 /(4m)$ for the longitudinal $\rho,\omega$ and zero otherwise.  For conventional reason, $4\pi^2$ will be used for $\phi$ meson instead of $8 \pi^2$.

The Borel sum rule with the  $\delta(u^2)$ ambiguity subtracted out is
obtained by taking the Borel transform of $\omega^2 \Pi^1(\omega)$.  This
gives,

\begin{eqnarray}
\left( m_\rho^2 f+ F(1-\frac{m_\rho^2}{M^2}) a \right)
e^{-m_\rho^2/M^2}-c_0 S_0 e^{-S_0/M^2} s  \nonumber \\ 
= 8 \pi^2 n_n 
\left( \frac{b_2}{M^2} -\frac{b_3}{2M^4} -b_{scatt} \right).
\label{borel2}
\end{eqnarray}
The $Q^2$ dependence in the $b$'s coming from the twist-2 operators changes to 
the 
 $M^2$ dependence\cite{SVZ}.  The vacuum parameters are first determined from 
the 
vacuum sum rules with parameters given as in ref.\cite{HL92}.  This gives
 $(m^2_V,F,S_0)= ( 0.77^2,1.48,1.43)$GeV$^2$ for the $\rho,\omega$ and 
 $(m^2_\phi,F,S_0)= ( 1.02^2,2.19,2.1)$GeV$^2$ for the $\phi$.  Then the parameters are determined by least square fit
 method.  The Borel interval for the transverse $\rho$ meson is determined by 
requiring that the contribution from dim 6 operators are less than 35\%  of
 the   dim 4 contributions, which determines the $M^2_{min}
\sim 1$GeV$^2$.  The maximum Borel mass is determined by requiring that the 
continuum contribution is again less than 35\% of the first term in the OPE,
which gives $M^2_{max} \sim 2.3 $GeV$^2$.  But it turns out that even if we 
choose higher $M^2_{max}$, the  result differ by less than 10\%.  The 
uncertainty  here comes from the less reliable number of the $b_3^t(\tau=4)$,
which gives less than 10 \%  uncertainty in the final answer.  
Overall, combining the uncertainty
coming from the continuum, we expect 40\% uncertainty for the numbers for 
 the transverse $\rho$ and also for the transverse $\omega, \phi$, which can be 
analysed in a similar fashion.  
 For the longitudinal directions of $\rho,\omega$, the dim 4 operators are suppressed by $\alpha_s$ and
the dim 6 operators contribute at tree level.  So  here we choose the 
 $M_{min} \sim 2.5$GeV$^2$, at which the dim 4 and dim 6 contributions are of 
similar order
 and take $ M_{max} \sim 3.5$ to $5$GeV$^2$.  Here, we expect 50 \% uncertainty.
 Similar analysis can be done for the longitudinal $\phi$.   
The results are shown in table 2, the numbers are all at the nuclear matter 
density.  For higher density, one can just multiply the numbers with the 
relevant ratio to the nuclear matter density.  Our results are consistent with 
effective hadronic model calculations.  First, as can be seen from table 2, the 
scattering term decreases the longitudinal $a$ value.  This is consistent with 
the  known result from the  nucleon-hole contribution \cite{Walecka}.  Second,  
recent calculation using resonance-nucleon hole 
contribution for the transverse part shows an attraction\cite{ben1}.  This is
also consistent with our $a$ values for the transverse part of the $\rho,\omega$.  

 As discussed before, a non-vanishing $a$ will shift the average 
 peak position by 
 $\Delta M= \sqrt{m_V^2+a {\bf q}^2}-m_V$, even if there is no change in the
scalar mass $m_V$.  With the values of $a$ obtained, we have plotted the 
fractional change $\Delta M/m_V$  for the $\rho$ meson in Fig. 1 a).  The solid 
lines denote the result at nuclear matter density, and the dashed lines 
that at 3 times nuclear matter density. The results for the $\omega$ meson 
look similar to Fig.1 a).  Fig 1 b). shows the result for the 
$\phi$ meson.  
As can be seen from the figures,   at nuclear matter
density, the expected shift is small, and will not wash out any shift of the 
peak position expected from the change of the scalar mass\cite{HL92}.  But, 
 at higher densities and higher ${\bf q} \sim 1$GeV, its effect becomes 
quite large.   
In  the dilepton production, the relative contribution
from the transverse and the longitudinal direction depends on 
the angle between the
total and relative three momentum of the outcoming dileptons\cite{GK91}.  
However,
the transverse direction has always a larger contribution.  In addition,
our result show that the residue $(F+f {\bf q}^2)$  increases more for the 
transverse direction
as ${\bf q}$ increases.   Therefore, the fact that $a <0$ for the 
transverse direction will shift the dilepton spectrum downwards for finite
 ${\bf q}$.   In addition, the contribution of the $\tau=2$ matrix element
is the same for the axial vector axial vector correlation function as in 
the case of vector and vector.  So the dominating effect that pushes 
 $a$ negative has nothing to do with chiral symmetry breaking and its
 restoration.  
 It is not clear yet, how much this effect will be in an actual RHIC.  
For that, one has to do a model calculation and such kind of calculation
is important to really pin down the possible change of vector meson mass
at finite density.   

We would like to thank Bengt Friman for introducing me to the importance 
of finite ${\bf q}$ effect, which initiated this study.  We thank T. Hatsuda and
H.C. Kim for useful comments.  This work was 
supported in part by the Korean
Ministry of Education through grant no. BSRI-96-2425, by the KOSEF through
grant no. 971-0204-017-2 and through the CTP at Seoul National 
University.

\begin{table}
Table 1. $b_2(q)$ ($b_2(G)$) represents the contribution of quark (Gluon) 
operators to $b_2$.    $b_{32}$ ($b_{34}$) represents the contribution of the 
$\tau=2$  ($\tau=4$) operators to $b_3$. The units for $b_2,b_3$ are
GeV and GeV$^3$ respectively
\vspace{0.5cm}
\begin{tabular}{ c   c c c c c c c c } 
 & $b_2^L(q)$ &  $b_2^L(G)$ & $ b_{32}^L$  & $b_{34}^L$   
 & $b_2^T(q)$ &  $b_2^T(G)$ & $ b_{32}^T$  & $b_{34}^T $   
 \\ \hline 
$\rho$ &-0.021 &-0.014 & 0.047  & 0.096 & 0.417 & -0.025 & 0.242 & 0.060 \\
$\omega$ & -0.021  &-0.014 & 0.047  & 0.096 & 0.417 & -0.025 & 0.242 & 0.138
\\
 $\phi$ & -0.004 & -0.028 & -0.003 & 0.052 & 0.074 & -0.049 & -0.002 &
0.015
\end{tabular}
\vspace{0.5cm}
\end{table}

\begin{table}
Table 2.   Results for the parameters at nuclear matter density.  The
values are from best fit of the Borel sum rule in eq.(\ref{borel2}).
The values in the bracket are the results without the scattering term.
\vspace{0.5cm}
\begin{tabular}{ c   c c c  c  } 
 & $a$ & $f$ & $s$ &
 Borel Interval GeV$^2$ \\ \hline 
Transverse $\rho$ & -0.065 & 0.137 & -0.008 & 
 1 $\sim$ 2.3
\\
Transverse $\omega$ & -0.040 & 0.120 & 0.009 & 
 1.3  $\sim$ 2.5 \\
Longitudinal $\rho,\omega$ & 0.021 & 0.068 & 0.027 
 & 2.5 $\sim$ 3.5 \\
  & (0.061)  & (-0.042) & (0.042) & \\
Transverse $\phi$ & 0.004 & 0.010 & 0.009 & 
 0.9 $\sim$ 2.0 \\
Longitudinal $\phi$ & 0.009 & -0.001 &
0.009 & 2.0 $\sim$ 3.0 \\
\end{tabular}
\vspace{0.5cm}
\end{table}

\begin{figure}
\caption{ a) The fractional change $\Delta M/m_V$ of the peak position of the 
$\rho$ as a function of ${\bf q}$.   The solid (dashed) lines show the results at nuclear matter (three times nuclear matter) density.  The positive changes 
correspond to the longitudinal  direction, the negative changes correspond the transverse directions.  
 b) The fractional changes of the $\phi$  as a function of ${\bf q}$.   The 
solid (dashed) lines show the result at nuclear matter (three times nuclear 
matter) density.  The larger changes correspond to the longitudinal  direction, the smaller changes correspond the transverse directions.  }
\label{fig1}
\end{figure}

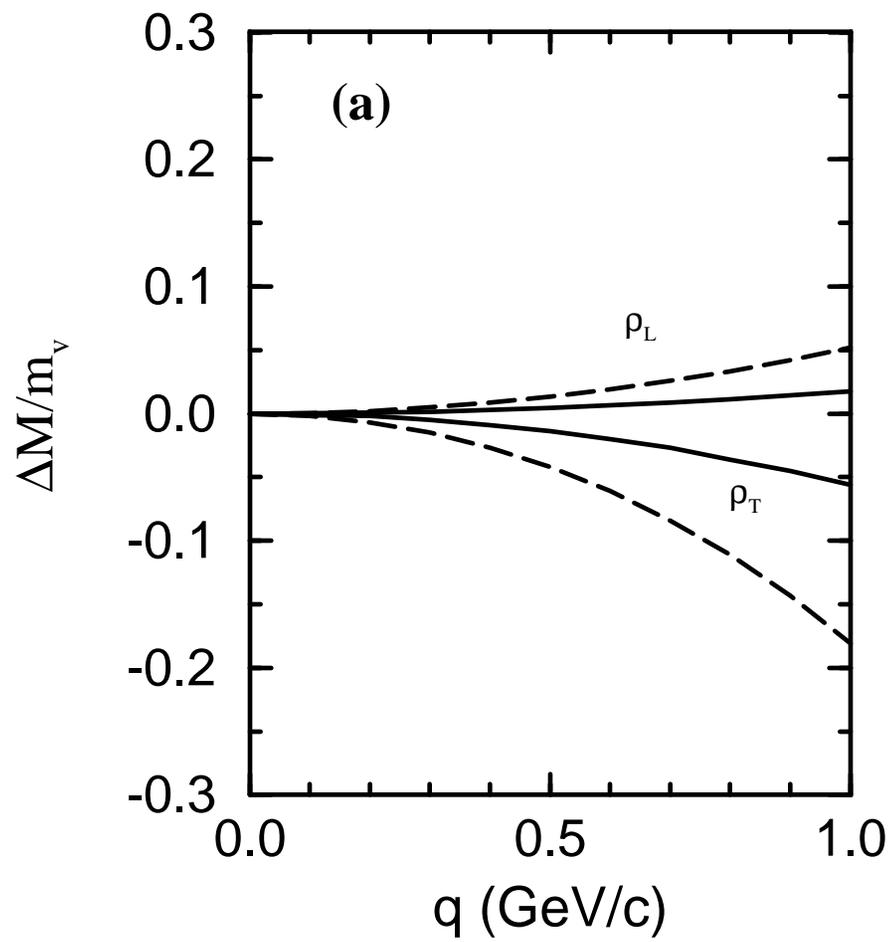
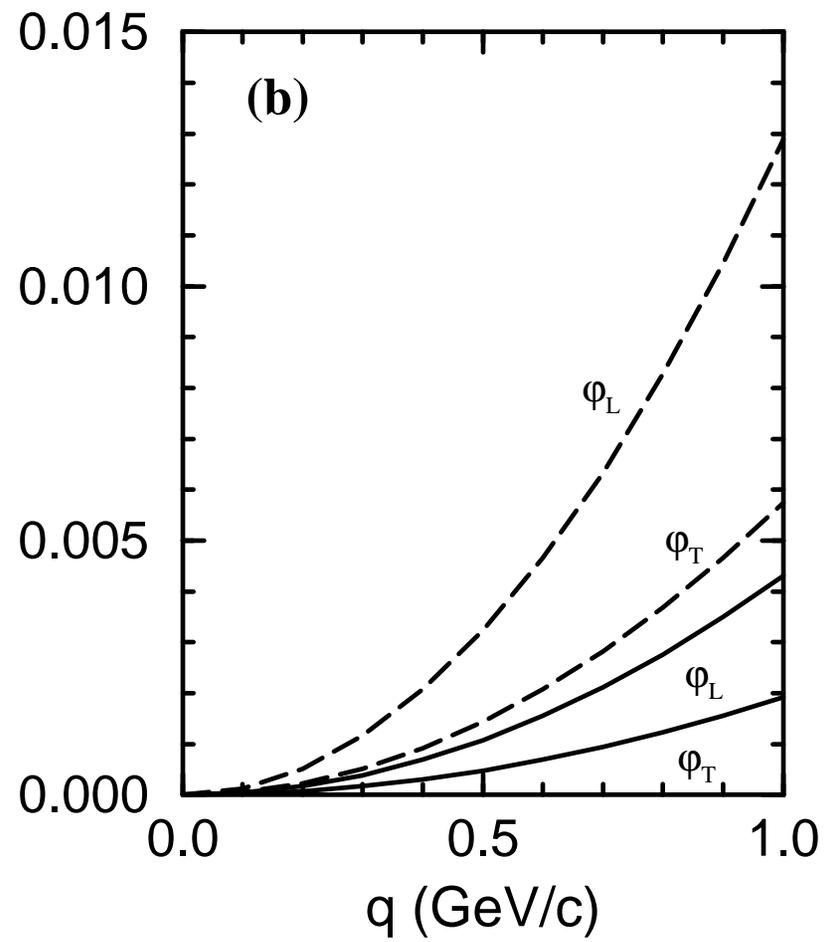
\end{document}